%% file: main.tex
\documentclass{article}

\usepackage{arxiv}
\usepackage[utf8]{inputenc} % allow utf-8 input
\usepackage[T1]{fontenc}    % use 8-bit T1 fonts
\usepackage{hyperref}       % hyperlinks
\usepackage{url}            % simple URL typesetting
\usepackage{booktabs}       % professional-quality tables
\usepackage{amsfonts}       % blackboard math symbols
\usepackage{nicefrac}       % compact symbols for 1/2, etc.
\usepackage{microtype}      % microtypography
\usepackage{lipsum}
\usepackage{hyperref}
\usepackage{tikzscale}
\usepackage{custom_aliases}
\usepackage{tikz}
\usetikzlibrary{backgrounds,patterns,decorations,positioning}

\def\BibTeX{{\rm B\kern-.05em{\sc i\kern-.025em b}\kern-.08emT\kern-.1667em\lower.7ex\hbox{E}\kern-.125emX}}

\title{Embedding Models for Recommendation under Contextual Constraints}
\author{Syrine Krichene \\ Criteo AI Lab \\
\href{mailto:s.krichene@criteo.com}{s.krichene@criteo.com}
\And Mike Gartrell \\ Criteo AI Lab \\ 
\href{mailto:m.gartrell@criteo.com}{m.gartrell@criteo.com}
\And Clement Calauzenes \\ Criteo AI Lab \\
\href{mailto:c.calauzenes@criteo.com}{c.calauzenes@criteo.com}
}

\begin{document}
\maketitle

\input{abstract.tex}
\input{introduction.tex}
\input{related-work.tex}
\input{controlled-embeddings-model.tex}
\input{experiments.tex}
\input{conclusion.tex}

\bibliographystyle{unsrt}
\bibliography{library}
\end{document}

%% file: abstract.tex
\begin{abstract}

Embedding models, which learn latent representations of users and items based on user-item interaction patterns, are a key component of recommendation systems. In many applications, contextual constraints need to be applied to refine recommendations, e.g. when a user specifies a price range or product category filter.
The conventional approach, for both context-aware and standard models, is to retrieve items and apply the constraints as independent operations. The order in which these two steps are executed can induce significant problems.  For example, applying constraints \emph{a posteriori} can result in incomplete recommendations or low-quality results for the tail of the distribution (i.e., less popular items). As a result, the additional information that the constraint brings about user intent may not be accurately captured.

In this paper we propose integrating the information provided by the contextual constraint into the similarity computation, by merging constraint application and retrieval into one operation in the embedding space.  This technique allows us to generate high-quality recommendations for the specified constraint. Our approach learns constraints representations jointly with the user and item embeddings. 
We incorporate our methods into a matrix factorization model, and perform an experimental evaluation on one internal and two real-world datasets. 
Our results show significant improvements in predictive performance compared to context-aware and standard models.
\end{abstract}

%% file: introduction.tex
\section{Introduction}
\label{sec:intro}
Embedding models, such as those based on matrix factorization, have become a core component of recommendation models. These models involve learning latent representations of users and items based on user-item interactions patterns, such as explicit feedback in the form of users' ratings on items, or as implicit feedback in the form of clicks. Many applications use contextual constraints to refine recommendations; for instance, a user may specify a price range or product category of interest. The contextual constraints can be seen as filters where the displayed recommendation have to fit the chosen filters.
The standard approach is to implement the retrieval of recommended items and the application of constraints as independent operations, which can lead to significant problems. Applying constraints either before or after retrieval can result in incomplete recommendations, or low-quality results for the tail of the distribution (less popular items). 

The constraints can be incorporated to the model as context features in a context-based model. Using those models, suppose that the context is correlated to the user and give as a static information about it. In other words one user can be described by the set of contexts that he choose in the past. In this paper we suppose that the setting may change and one user can choose any context independently of his history but the constraint distribution still constant. This choice of setting is motivated by the sparsity of the observed data (the contextual constraints) and the characteristics of the constraints that can be continuous. The new context representation can be orthogonal to the user's context representation and this can lead to spurious recommendation.

Instead of treating retrieval and constraint application as separate, independent operations, we propose merging filtering and retrieval into one operation in the embedding space by integrating information provided by the constraint into the similarity computation. We implement this by using the constraint to modify the user's embedding vector, since the constraint specifies contextual information that denotes a similarity between the user and a class of items (that verifies the context). We present several different parametrizations of constraints, suitable for category-based filters. For constraints based on a continuous range of values (e.g., price, duration, etc.) we discretize and extract categories.  User-item interactions for some types of contexts or constraints may be quite sparse. In this paper we focus on contextual constraints described by a set of features and not by words we then exclude search and more generally natural language processing tasks.

Our method consist of representing the context constraints as a transformation of the embedding space. We still learn one vector representation for each user and each item.

The main contributions of this work are the following:
\begin{itemize}
   \item We introduce a new framework that describes contextual constrained recommendation, using sparse data, and discuss the limitations of baseline models. 
   \item We propose a new approach adapted to a new setting for learning contextual constraint representations. The contextual constraint information incorporated into a matrix factorization model as a transformation of the user embedding space. 
   \item We introduce linear and non-linear transformations of the user embedding space and adapt these methods to neural matrix factorization models.
   \item We evaluate different configurations of our approach, on different settings. These settings are motivated by real-life use cases. We compare our models to context-aware and non-context-aware models on one private and two public datasets. 
\end{itemize}

We present the contextual setting framework and a review of related work and background on constraints for recommendation systems in Section~\ref{sec:related-work}.  In Section~\ref{sec:model} we introduce our modeling approaches for contextual constraints adapted to matrix factorisation model and neural matrix factorisation. Section~\ref{sec:experiments} shows that our approaches can lead to significant improvements in predictive performance compared to baseline approaches.

%% file: related-work.tex
\section{Framework and Related Work}
\label{sec:related-work}
\subsection{Framework for Recommendation under contextual constraints}
A common approach to recommendation is to retrieve items $i \in [n]$ for a user $u \in [m]$ depending on the items' relevance to the user. The relevance of an item to a user is only observed a posteriori, once the user has purchased or interacted with the item and provided feedback $r$; we denote the matrix containing feedback $r$ for each pair $(u,i)$ as $R \in \{0,1\}^{m \times n}$. The vast majority of the entries in $R$ are unobserved, since users provide only limited feedback.

Hence, in order to recommend new items to the user, one needs to model the conditional distribution of $r$ given $(u,i)$ to infer the unobserved entries of the feedback matrix $R$. While $R$ may contain different types of feedback (such as ratings, or the number of interactions), up to a normalization, we assume that $r \in [0,1]$. Practically, a simple approach is to model the conditional expectation of $r$, which can be seen as a similarity between $u$ and $i$:
\begin{align*}
    s_{u,i} \triangleq \lE[r | u, i] =  \argmin_{s} \lE\left[ \left(R_{u,i} - s\right)^2 \right] \,.
\end{align*}
Then, we can fit a parametric modelling $\hat{s}_{u,i}$ of $s_{u,i}$ by minimizing the empirical squared loss\footnote{We consider the squared loss for simplicity, but it could be replaced by any other Bregman divergence $l$ (ex: log-loss), as they all satisfy $\lE(Y|X=x) = \argmin_x \lE_Y(l(Y,X)|X=x)$.} on the data $D_t$ available at training time:
\begin{align}
     \min_{\hat{s}_{u,i}} \frac{1}{t}\sum_{(u,i) \in D_t}  \left(R_{u,i} - \hat{s}_{u,i}\right)^2
\label{eq:empirical_squared_loss}
\end{align}
where $t$ is the number of training instances. We refer the reader to Sec. \ref{ssec:related_work} for common modelling of $\hat{s}_{u,i}$.

\paragraph{Constrained Retrieval}
In this work, we want to take into account how the feedback $r$ depends on some constraints on the items $i$ rated by the user $u$, which should be treated as a context as soon as these constraints are informative of the feedback distribution. We want to take into account constraints of the form: "items of brand X", "restaurants opened at 7pm", "a movie in either the action or comedy genre", or disjunctions of the latter. To formalize this, we assume that there exists a \emph{feature} map $f : [n] \to \{0,1\}^{d}$ that associates one or several values in a finite set $[d]$ to each item $i$, e.g. the genres of a movie. A set of values is represented by a binary vector in $\{0,1\}^d$. For  simplicity of notation, we will denote $f(i)$ as $f_i$. Finally, as a constraint $c$ can be a disjunction of values, it can also be formalized as the elements of the power set of $[d]$, represented by binary vectors $c\in \cC \subseteq \{0,1\}^d$. An item $i$ is said to satisfy the constraint if $c^T f_i > 0$, i.e. if the item $i$ has at least one value of the feature in common with the constraint $c$ (e.g. one matching movie genre).

\paragraph{Learning with constraints:} Now that we have formalized the set of constraints $c$ that can be input by a user $u$ to restrict the set of items she aims for, we consider our modeling of the reward process. We need to model the conditional distribution of $r$ given $(u,i,c)$, and we assume that we have a 3-D tensor $R \in \lR^{m \times n \times |\cC|}$ of feedback. Ultimately, we are interested in estimating the conditional expectation:
\begin{align}
     s_{u,i,c} \triangleq \lE[r | u,i,c] = \argmin_{s} \lE\left[ \left(R_{u,i,c} - s\right)^2 \right]
\label{eq:constraints-squared-loss}
\end{align}

 It is clear that due to the combinatorial nature of $\cC$, we need to carefully consider our modeling choice for $\hat{s}_{u,i,c}$, in order to keep the inference complexity manageable and avoid substantial estimation problems.
Given a finite set of available data $D_t$, we will fit $\hat{s}_{u,i,c}$ by minimizing:

\begin{align}
     \min_{\hat{s}_{u,i,c}} \frac{1}{t}\sum_{(u,i,c) \in D_t}  \left(R_{u,i,c} - \hat{s}_{u,i,c}\right)^2
\label{eq:empirical_contextual_squared_loss}
\end{align}

Note here that we implicitly assume the joint distribution of $(u,i,c)$ to be the same at training and test time, i.e., there is no distribution shift. A more robust (and statistically harder) task would be to learn to predict in the worst case over the possible constraints that the user may input at test time. Formally, this could be handled by replacing the marginalization over $c$ in \eqref{eq:empirical_contextual_squared_loss} by an infinite norm. However, this norm would need to be relaxed, since the infinite norm is not straightforward to optimize; we therefore omit further discussion of the infinite norm.

\subsubsection{Causal Graphical model}
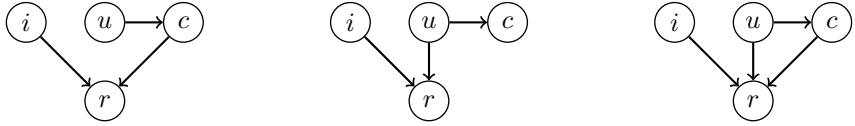
\begin{figure}[!ht]
    \centering
    \begin{tikzpicture}[auto, thick, node distance=0.5cm]
    \tikzstyle{var_node}=[circle,draw,minimum size=15pt,inner sep=0pt, line width=0.5pt]
    \node[var_node] (user) {$u$};
    \node[var_node, left=5mm of user] (item) {$i$};
    \node[var_node, right=5mm of user] (filter) {$c$};
    \node[var_node, below=5mm of user] (reward) {$r$};
    \draw[->] (filter) -- (reward);
    \draw[->] (user) -- (filter);
    \draw[->] (item) -- (reward);
    \end{tikzpicture}
    \hspace{1.5cm}
    \begin{tikzpicture}[auto, thick, node distance=0.5cm]
    \tikzstyle{var_node}=[circle,draw,minimum size=15pt,inner sep=0pt, line width=0.5pt]
    \node[var_node] (user) {$u$};
    \node[var_node, left=5mm of user] (item) {$i$};
    \node[var_node, right=5mm of user] (filter) {$c$};
    \node[var_node, below=5mm of user] (reward) {$r$};
    \draw[->] (user) -- (reward);
    \draw[->] (user) -- (filter);
    \draw[->] (item) -- (reward);
    \end{tikzpicture}
    \hspace{1.5cm}
    \begin{tikzpicture}[auto, thick, node distance=0.5cm]
    \tikzstyle{var_node}=[circle,draw,minimum size=15pt,inner sep=0pt, line width=0.5pt]
    \node[var_node] (user) {$u$};
    \node[var_node, left=5mm of user] (item) {$i$};
    \node[var_node, right=5mm of user] (filter) {$c$};
    \node[var_node, below=5mm of user] (reward) {$r$};
    \draw[->] (user) -- (filter);
    \draw[->] (user) -- (reward);
    \draw[->] (filter) -- (reward);
    \draw[->] (item) -- (reward);
    \end{tikzpicture}
    \caption{Causal graphical models for different settings while observing contextual constraints $c$ for a user $u$, who interacted with an item $i$, and gave feedback $r$.}
    \label{fig:proba_graph}
\end{figure}

The causal graphical model of a dataset is the causal representation of the variables of the generative process of the data. In our setting we have the item, the user, the rating, and the constraint. There are several different possible representations for a given dataset. Usually we try to simplify the graph, and use the simplest possible model. Figure~\ref{fig:proba_graph} shows the different possible causal graphical models for our setting (left to right):
\begin{itemize}
    \item In the first graph, we suppose that the user doesn't directly cause the final rating, and it is only through the constraint that he impacts the rating for an item. Therefore, conditioning on both the item and the constraint allows us to fully explain the rating. The best model to use in this setting could be a simple linear model, with enough features to fully describe the constraints.  This setting requires a complete description of the constraints.
    \item In the second setting, we suppose that the constraint doesn't have a direct impact on the rating. However, the constraint is correlated with the rating: the constraint indirectly informs us about the unknown user intent $u$, and thus conditioning on it still allows us to improve the rating prediction.
    \item The third setting is simply the combination of both previous scenarios. 
\end{itemize}
While in theory the first and third settings are perfectly valid, the reality lies more often in the second setting: observing the constraint informs us about the user's hidden intent, allowing to better predict the rating. 

\subsubsection{Contextual constraints provide information about instantaneous user utility}
In cases where we observe the second graph in Figure~\ref{fig:proba_graph}, we know that the contextual constraint cannot impact the reward, as there is no causal link between the two nodes. However, knowing the contextual constraints set by the user may help us to refine the model. The constraints give us additional information about the intent of the user, and the observed reward varies according to the user intent, which is proportional to the user's instantaneous utility. ~\cite{loewenstein2003time} describes the importance of time on the instantaneous utility of an item for a user: The instantaneous utility is usually described as a smooth convex function, which gives the reward for an item as a function of time. The utility time corresponds to the elapsed time from the moment when the user first observed the item. The utility estimates the extent of the user's interest in purchasing an item, while the reward is the feedback from the user.  The reward can be observed after the purchase, in case of item ratings, or immediately after the recommendation, such as for the click-based recommendation task. In case of ratings, it is possible to observe a maximum utility followed by a low reward: if the user is not satisfied with the purchased item, then we expect a very low reward (since the user had significant interest in purchasing the item, and is then disappointed with the item after purchase). For both cases, the reward is at its optimum when the instantaneous utility's maximum is reached. Therefore the observed reward may depend on the time at which it is observed.

The instantaneous utility may introduce a bias in the learned model. For example, suppose we learn a model that predicts the probability of click under a normal setting, with no observed constraints. The model may recommend a good item for a particular user, but if the instantaneous utility is low, then the user may not click. In this case our evaluation metrics will indicate that the behaviour of the model is incorrect, and this will increase the variance of the evaluation metric. Another use case we consider occurs when the observations used for training contain non-clicked items observed with a very low user utility. In this case, the model will learn that such items are very bad to recommend for any user, which introduces error in the prediction.

In general recommendation settings, the dataset often doesn't provide any information about instantaneous utility for users. In the contextual constraint-based setting (where all the observations contain contextual constraints), by setting a constraint the user gives us information about her intent, indicating that she is looking for an item with specific characteristics. Given this information we suppose that the expectation of the instantaneous utility computed over all users for the constraint setting is higher than the expectation computed for the general setting without constraints. We also expect lower variance for the contextual constraint setting.  These phenomena reduce the prediction and learning error. In other words, the second and third graphs shown in Figure~\ref{fig:proba_graph} can benefit from the contextual constraint setting when learning a contextual or non-contextual model.

\subsection{Background and related work}
\label{ssec:related_work}

Matrix factorization (MF) models have been one of the most popular and successful recommendation approaches of the past decade~\cite{koren2009matrix, mnih2008probabilistic, salakhutdinov2008bayesian, jamali2010matrix, ma2008sorec, gopalan2015scalable, chaney2015probabilistic, paquet2013one, stern2009matchbox, kula2015metadata, hu2008collaborative, rennie2005fast, srebro2005maximum, pan2008one, rendle2009bpr, gantner2012personalized, pan2009mind, su2009survey} to model $\hat{s}$. These models learn latent representations of users and items based on patterns in observed interactions between users and items, using loss functions that are similar to \eqref{eq:empirical_squared_loss}.  Many variants of MF models exist, including probabilistic models~\cite{mnih2008probabilistic, salakhutdinov2008bayesian, gopalan2015scalable, chaney2015probabilistic, paquet2013one, stern2009matchbox}, models for implicit feedback~\cite{gopalan2015scalable, paquet2013one, hu2008collaborative, pan2008one, rendle2009bpr, pan2009mind}, models that support metadata~\cite{stern2009matchbox, kula2015metadata}, and models that incorporate social network information~\cite{jamali2010matrix, ma2008sorec, chaney2015probabilistic}, among others.

\subsubsection{Matrix Factorization}
The first family of approaches that we consider are MF models that do not involve context. MF models are based on dimensionality reduction. Let $U, P$ represent the embedding matrices for users and items, respectively, and $k$ represent the number of embedding dimensions. We therefore have $U \in \lR^{n\times k}$ and $P \in \lR^{m\times k}$, and the similarity function $\hat{s}$ between $U$ and $P$ is given by:
\begin{align}
    \hat{s}_{u,i} = U_u P_i^T + B_u
\end{align}
where $B \in R^{n}$ is the bias term for the user. We fit $U,P,B$ by minimizing a penalized version of \eqref{eq:empirical_squared_loss}:
\begin{align}
    \min_{U,P,B} \sum_{(u,i)\in D_t} \Big( R_{u,i} - \underbrace{\left(U_u P_i^T + B_u\right)}_{\hat{s}_{u,i}} \Big)^2 + \frac{\lambda}{2} \left(||U||_2^2 + ||P||_2^2 \right)
\end{align}
This model supposes that we ignore context, or gain no additional information from context, and the similarity between items and users is completely captured by examining the ratings. At recommendation time, cosine similarity is often used to compute the $k$ best products to recommend to a specified user. Cosine similarity is computed over the vector representations of the user and the item. If the user and item vectors are orthogonal then the user is far from the item, and therefore the item will not be recommended to this user. If the user and item vectors have the same orientation, then the cosine similarity is at it's highest value of 1, and the item will be recommended for this user. The $k$ best products to recommend are those that maximize the cosine similarity score.

\subsubsection{Context-aware Matrix Factorization}
Context-aware recommender systems (CARS) have been proposed in order to address a limitation of conventional recommendation models and incorporate contextual features into these models. In CARS, contextual features can be related to users or items. We consider the contextual constraints as item features, as context may change the utility observed for a particular item. This allows us to use Context-Aware Matrix Factorization (CAMF-CI), presented in~\cite{baltrunas2011matrix}, as a baseline model. If we consider contextual constraints as contextual features, then our loss function is:
\begin{align}
\min_{U,P,B,C}  \sum_{(u,i,c)\in D_t} & \Big(R_{u,i,c} - \underbrace{\left(B_u + U_u P_i^T + C_{c,i} \right)}_{\hat{s}_{u,i,c}} \Big)^2 + \nonumber \\ & \frac{\lambda}{2} \left(||U||_2^2 + ||P||_2^2 + ||C||_2^2 \right) 
\end{align}
where $C\in R^d\times R^m $ is the contextual feature matrix, and we have one variable to learn per feature per item. In practice we learn only the variable where the feature $f$ is compatible with the item $i$.
Due to the sparsity of the contextual constraint data the probability of observing orthogonal context vectors is high. If at recommendation time a user specifies a context with a vector that is orthogonal to his history, then the recommendation will be almost random since the model will recommend items from the tail of the item popularity distribution.

\subsubsection{Context-aware collaborative filtering MF: NNMF}
Context-aware collaborative filtering systems are a class of recommendation systems based on the past experiences of the user. Some of these models use the user's timeline of past item interactions to predict future preferences. In our setting the timeline is not given and we assume that we cannot recover it. Context-aware collaborative filtering models use multiple features that describe users and items, which is not the case in our setting, as we only have access to the user id, item id, and contextual features. For all collaborative filtering models,  features related to the user's past perfectly describes the user's feedback. In this paper we suppose that for a particular user, the set of contexts is not fully explored and may change at recommendation time. Therefore, it is problematic to recommend items to a user under a context vector that is orthogonal to the context vectors seen in his past. For example the model introduced in~\cite{karatzoglou2010multiverse} learns one representation of a user for each context. If the user have never been seen in a particular context, then contextual representation of that user will be random.
We use as a baseline model an adaptation of a neural matrix factorization model~\cite{volkovs2017dropoutnet}, where the embedding vector of a user or product is learned as the output of a neural net (NNMF):
\begin{align}
    \min_{U,P} \sum_{(u,i,c)\in D_t} & \Big( R_{u,i} - \underbrace{\left(U(u,c) P(i,c)^T\right)}_{\hat{s}_{u,i,c}} \Big)^2
    \label{eq:nnmf}
\end{align}
with $C$ is the output vector of the neural net, which takes as input the contextual feature $c$ and the item $i$. Unfortunately, the high capacity of this model is more of a handicap when the context space $\cC$ is combinatorial: it is very prone to over-fitting.

%% file: controlled-embeddings-model.tex
\section{Adapted Embedding Models for Contextual Constraints}
\label{sec:model}
On one hand, we aim for more expressivity than CARS models, as the context encodes constraints which represent information regarding the  user-item \emph{interaction}, and not simply an additional item-centric or user-centric component to the score. On the other hand, we can't afford to have as much flexibility as NNMF due to the combinatorial nature of $\cC$. To resolve this dilemma, we propose using the context $c$ to parameterize the similarity between the user and item in the MF embedding space.  Our approach represents the constraints as a transformation of the embedding space. Given two embedding matrices $U \in \lR^{m \times k}$ and $P \in \lR^{n \times k}$, and a transformation $T:\cC \to \lR^{k\times k}$, we have
\begin{align}
    \min_{U,P, T, B} \sum_{(u,i,c)\in D_t} & \Big( R_{u,i} - \underbrace{\left(U_u T(c) P_i^T + B_u\right)}_{\hat{s}_{u,i,c}} \Big)^2
    + \frac{\lambda}{2} \left(||U||_2^2 + ||P||_2^2 \right)
    \label{eq:loss_model}
\end{align}
This decomposition can be interpreted in two ways: $U_u T(c)$ represents the user in the context of the constraint provided as input, and $P_i T(c)^T$ represents an item in the context of a given filter. 
This formulation also has several useful scaling properties. First, the user embeddings $U$ and item embeddings $P$ are independent from the context, which allows us to avoid the need to parameterize the embedding computation as in NNMF, and help prevents overfitting. Second, contrary to CARS, this decomposition encodes the effect of the context on the \emph{interaction} between users and items. Lastly, with careful choice of $T$, the minimization show in \eqref{eq:loss_model} can be done in an alternating fashion, which has several advantages compared to stochastic gradient descent of the joint loss in~\eqref{eq:loss_model}, such as straightforward parallelization and better convergence properties. 

For our experiments we use several different formulations of $T$. If we look at ~\eqref{eq:loss_model}, while updating the user vector representation, the model will learn only one representation for the user that satisfies different contexts. This task is more difficult than the standard MF task. However, due to the sparsity of the data, a user rarely sets a very high number of contexts, and the task then becomes easier to solve.

\subsection{\texorpdfstring{$T$}{T} as a linear transformation}
Our first approach for formulating $T$ is to define it as a linear operator over a set of base transformations. Denoting $A \in \lR^{k \times k \times d}$ as a tensor containing a $k$-by-$k$ matrix for each value in $[d]$, we define
$$ \forall c \in \cC, \,\,\,\,\,T(c) = \frac{Ac}{||c||_1} $$
Depending on the size of $k$, we can use relatively simple variants of $A$. For example, we could use a low-rank factorization of $A$, or simply a slice-wise diagonal structure. For the sake of scalability, in our experiments we choose to constrain $A$ to be diagonal for each slice, i.e., for any $j \in [d]$, the $k$-by-$k$ matrix $\left[ A\right]_{j}$ is diagonal. We denote the resulting algorithm, using this formulation of $T$ and optimizing \eqref{eq:loss_model}, as \emph{diagonal constraints for matrix factorization}, or \textbf{DC-MF}. In Section~\ref{sec:experiments}, we also use a more constrained version as a baseline, which corresponds to setting $\alpha_1, \dots, \alpha_d \in \lR$ such that for any $j\in[d]$, $\left[ A\right]_{j} = \alpha_j I_k$, where $I_k$ denotes the identity matrix of size $k$. We denote this version of the model as \emph{weighted constraints for matrix factorization} (\textbf{WC-MF}). We expect this model to provide good performance for settings with a high number of overlapping constraints, since it extracts correlations between features.

Regarding DC-MF, one may find the choice of the diagonal structure to be too restrictive. However when $d$, the number of different values spanned by the feature map $f$, is lower than the embedding size $k$, this formulation usually proves to be sufficient. For cases where $d$ is very large compared to the embedding size $k$, there are different options. If no additional information is available, then we could use the full tensor $A$, instead of a diagonally constrained tensor. However, when $d$ very large, it is generally the case that utilizing additional information to describe the constraints would be very useful. In the next section, we propose a way to incorporate into $T$ such additional information for describing the constraint $c$, using a neural network.

\subsection{\texorpdfstring{$T$}{T}  as a nonlinear transformation}
When the space of constraints is very large, using additional information to describe the constraint $c$ allows us to keep the complexity manageable. For example, in the case of movie recommendation, there may be rules such as "do not recommend horror movies to a child under 10 years old". This type of constraint can be described in a richer way than is possible with a binary vector encoding "not horror". Similarly, when the user can specify constraints based on several features (for example, color, size, brand), it is more efficient to use a richer representation, rather than simply the Cartesian product of the features.

We formalize this by a \emph{constraint feature map} $g : \{0,1\}^d \to \lR^p$, which associates some additional information $g(c)$ with the description of a constraint $c$. We then define $T$ in a parametric way, for instance using a neural network $h_\theta : \lR^p \to \lR^{k \times k}$ (with parameters $\theta$). $T$ is therefore defined as:
$$
\forall c\in\cC \,\,\,\,T(c) = h_\theta(g(c))
$$
and $\theta$ is fit in the joint optimization of~\eqref{eq:loss_model}. We refer to the resulting model as \emph{neural constraints for matrix factorization} (\textbf{NC-MF}). It is particularly interesting to compare this model to the baseline NNMF, to determine if we manage to retain enough expressivity while sufficiently reducing the complexity of the network to avoid over-fitting. We perform this experimental comparison on the MovieLens dataset in Section~\ref{ssec:nn_experiments}. When comparing to the \textbf{NNMF} baseline it is preferable to use \textbf{NC-NNMF}, instead of \textbf{NC-MF}, to ensure a fairer comparison. The input features are the same for \textbf{NNMF} and \textbf{NC-NNMF}, although the features are used differently. In \textbf{NC-NNMF} the context features are used only for the transformation, while for \textbf{NNMF} they are used either as user input or item input features. 

\subsection{Extensions and Limitations}
\subsubsection{Constraints based on Multiple Features:} If we need to integrate several feature maps $f^{(1)}, \dots f
^{(k)}$ (for example, brand, color, and size for clothes) taking values in $\{0,1\}^{d_1}, \dots, \{0,1\}^{d_k}$, we can define the feature map $f$ to take values in the Cartesian product of the definition spaces of the different features $\prod_{j=1}^k \{0,1\}^{d_j}$ (which is still a finite set). In this case, constraints $c$ can be conjunctions (over the different features) of disjunctions (over the values of each features), which can be still defined as a point in the power set of $\prod_{j=1}^k \{0,1\}^{d_j}$. This directly fits our setting, and w.l.o.g, we can restrict ourselves to the case of \emph{one} feature only taking a \emph{finite} number of values for the sake of simplicity. However, the practitioner should instead choose the parameterization of $T$ with a neural network (NC-MF) to keep the complexity manageable.

\subsubsection{Constraints based on Continuous Features:} Sometimes the constraints are defined from real-valued feature maps. In our experiments we choose to handle such continuous features by discretizing them, as treating these features as continuous often requires a model that is specific to the nature of the information represented by the feature (see Section~\ref{sec:experiments} for further details).

%% file: experiments.tex
\section{Experiments}
\label{sec:experiments}
To validate the proposed approaches, we conduct experiments on one private dataset and two public datasets. We choose these three datasets as they illustrate different possible settings that fit within our framework. The settings present three different tasks to solve. For each task we will compare our methods (DC-MF and NC-MF) to baseline models (MF, CAMF-CI, NN-MF), and discuss the limitations of each model. Due to the small size of the data we use area under the curve (AUC) as a metric to evaluate the quality of the prediction, since per-user ranking metrics are not appropriate in this small-data setting.

\subsection{High number of overlapping constraints}
In this setting we assume that one item can be observed with different contextual constraints: given the observed item $i$ we observe multiple constraints $c\in\cC$ such that $c^T f_i>0$. Intuitively, some transfer learning between the constraints can be used to refine the recommendation.
For popular constraints, the MF model is expected to achieve good results, since popular constraints correspond to average behavior. Given two popular contexts, we expect to observe, with high probability, a non-empty intersection in the item space, i.e., $\sum_{i} \left( c_1^T f_i\right) \left( c_2^T f_i\right)  > 0$. For MF, at recommendation time there will be a low probability that the $k$ best items selected, which are compatible with $c_1$ or $c_2$, will reside in the tail of the predictive distribution.

To better model the behavior of the constraint distribution in the tail, an alternative model to use is contextual matrix factorization, where the context is presented as a translation term (CAMF-CI). Using the context will refine the recommendation, and we therefore expect better results than with MF. As the CAMF-CI model is still crude for combinatorial constraints (since it cannot perform transfer learning from one constraint to another), it is still hard to learn on non-popular contexts. At recommendation time, when two features are rarely set together, the prediction using the learned weights will be almost random, since the learned model didn't observe this particular combination of features.

In contrast, our models allow some transfer learning from one constraint $c_1$ to another $c_2$, when $c_1^T c_2 > 0$. This allows our models to generalize better in the tail of the constraint distribution. We show that our model has more expressivity than CAMF-CI, and more effectively handles the combinatorial nature of $\cC$. These new methods are more adapted to the sparsity of the data in our framework. In this experiment, we do not consider neural-network-based methods, since the dataset used (from \emph{Foursquare}) is too small. Instead, we compare neural network methods in Sec.~\ref{ssec:nn_experiments}.

To better learn the transformation $T$ in the DC-MF model, we use a warm start heuristic to initialize the $A$ tensor. Because of its diagonal structure, we represent $A$ with a $k$-by-$d$ matrix. We initialize $A$ such that the columns ${\left[ A \right]}_{j}$ and ${\left[A \right]}_{j'}$ corresponding to values $j, j'$ of the constraints that frequently co-occur, have nearly the same value. Therefore, we initialize $A$ by minimizing:
\begin{align}
     \min_{A} \frac{1}{t}\sum_{c \in D_t} (I_k - (A \odot c)^T(A \odot c))^2
\label{eq:warmstart-featuresimilarity}
\end{align}
We use this warm start heuristic to facilitate transfer learning when training the model.

\begin{figure}[htbp]
\centering
     \includegraphics[width=0.48\textwidth]{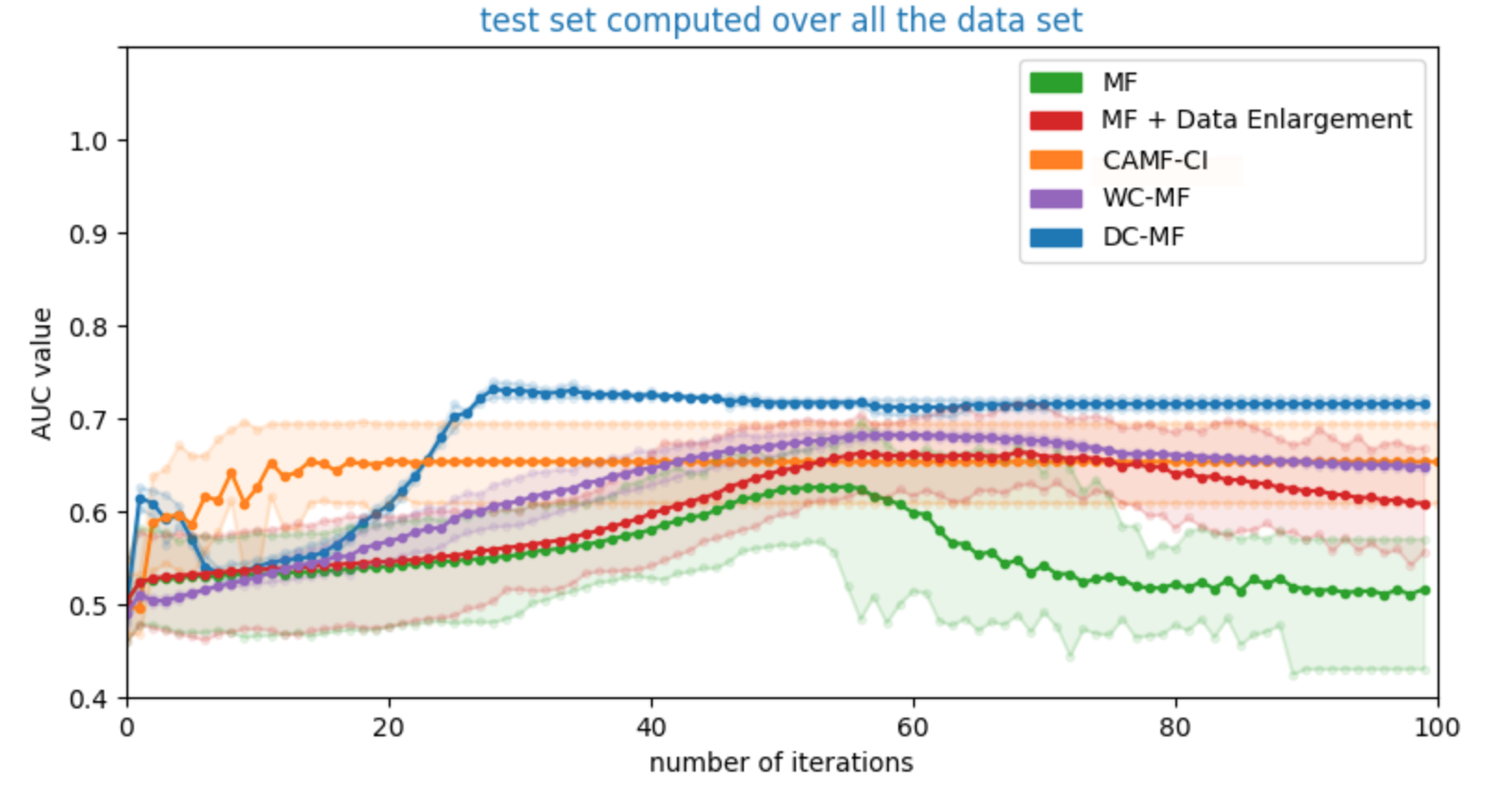}
    \caption{Comparing results for linear models evaluated on the Foursquare dataset.  Results show AUC on the global test set.}
    \label{fig:foursquare_all}
\end{figure}

\begin{figure}[htbp]
\centering
     \includegraphics[width=0.48\textwidth]{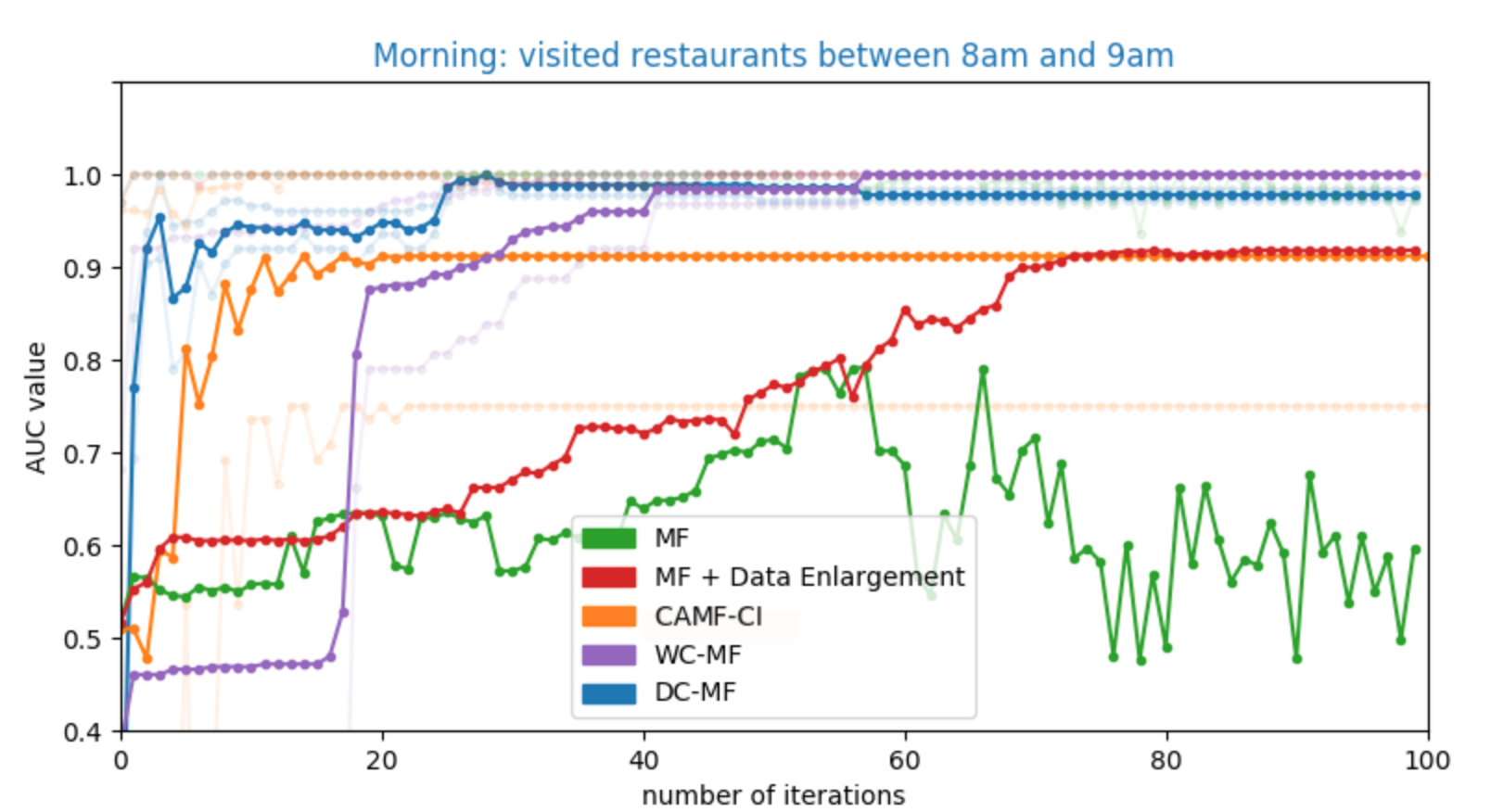}
    \caption{Reported AUC for Foursquare data for a rare context; the check-in time is between 8am and 9am.}
    \label{fig:foursquare_morning}
\end{figure}

\begin{figure}[htbp]
\centering
     \includegraphics[width=0.48\textwidth]{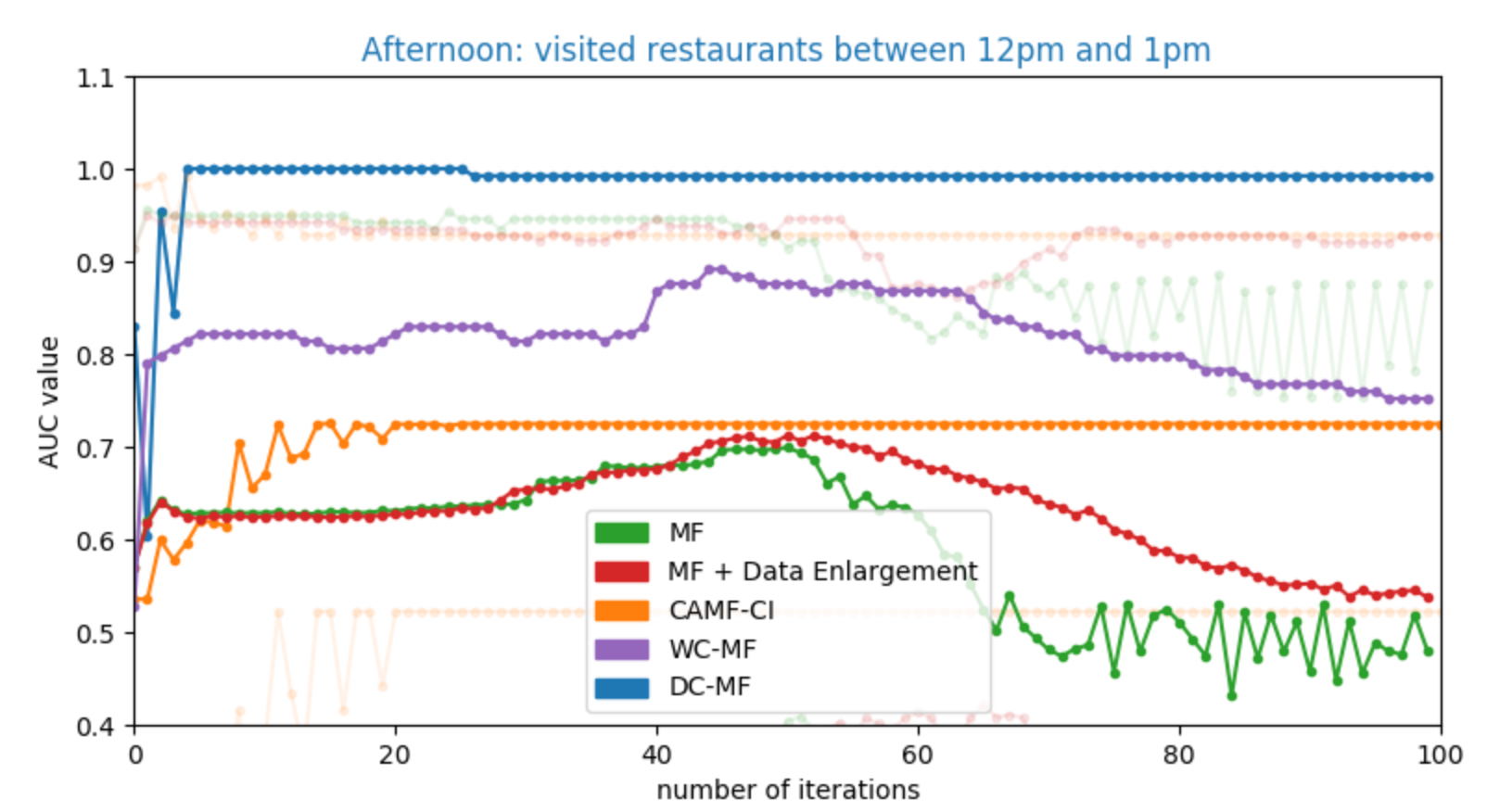}
    \caption{Reported AUC for Foursquare data for a rare context; the check-in time is between 12pm and 1pm.}
    \label{fig:foursquare_afternoon}
\end{figure}

\begin{figure}[htbp]
\centering
     \includegraphics[width=0.48\textwidth]{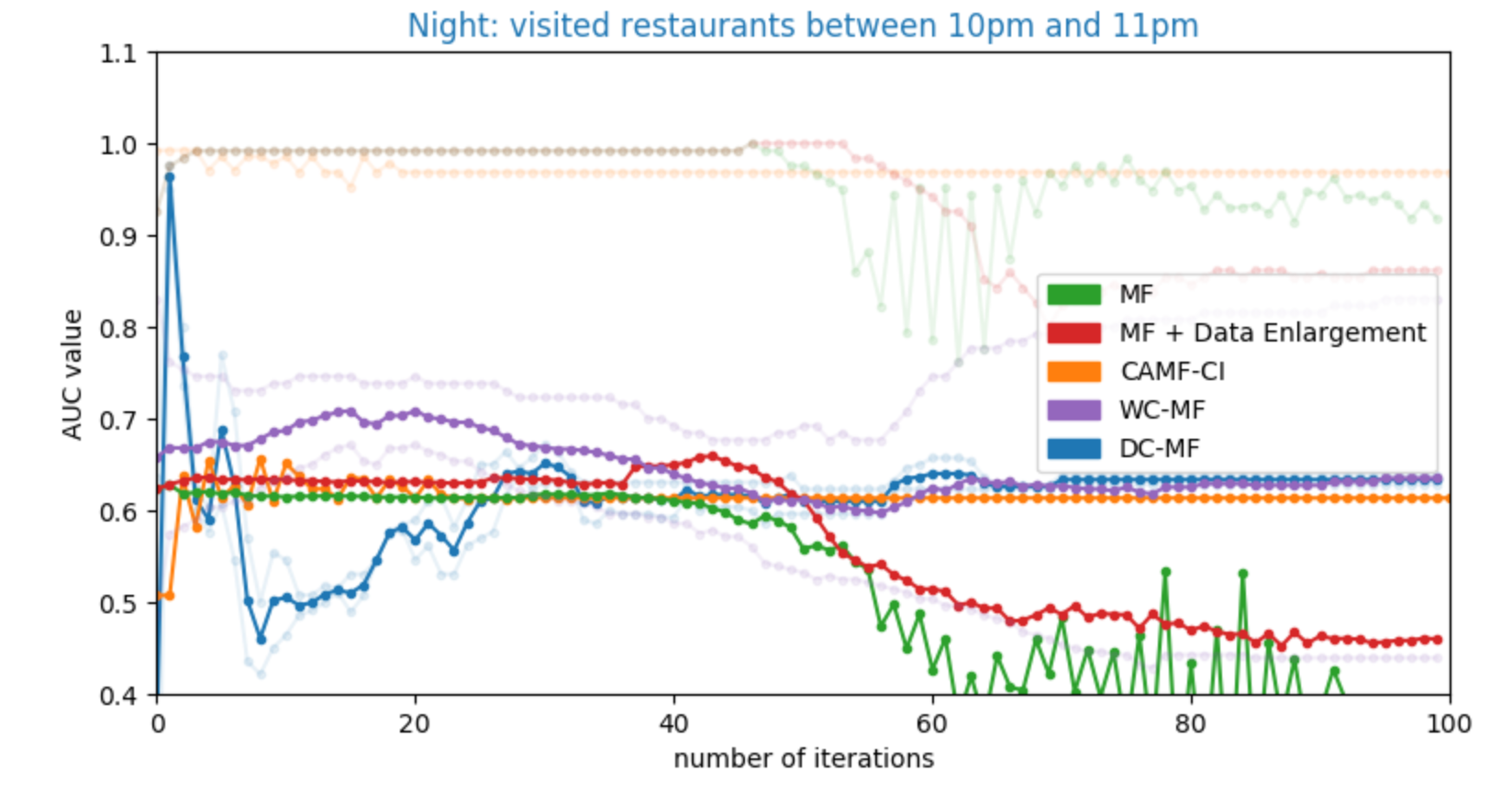}
    \caption{Reported AUC for Foursquare data for a popular context; the check-in time is between 10pm and 11pm.}
    \label{fig:foursquare_night}
\end{figure}

\paragraph{Foursquare dataset}
This Foursquare NYC check-in Dataset contains $227,428$ check-ins in New York city. Each check-in is associated with a time stamp, GPS coordinates, and a category for the venue. We leverage this dataset to study recommendations under time-based contextual constraints. We use only information about the user id, restaurant id, and check-in time in our experiments.
We assume the following:
\begin{itemize}
    \item A user only checks in to restaurants that he likes. Therefore, we observe only positives for this dataset.
    \item A user cannot visit a restaurant that is not open.
    \item A user visits different restaurants according to the type of meal that she is looking for (breakfast, lunch, or dinner).
\end{itemize}
Both the contextual constraints of opening hours and serving a particular meal type are time dependent. We can therefore assume that the check-in time represents a contextual constraint. We discretize time in order to represent it with binary features. As the exact constraints are approximated by the check-in time, we relax $c^T f_i$ for each restaurant $i$ to cover one hour (corresponding to five values in the discretization) around the observed check-in time. By construction, each restaurant $i$ visited by the user $u$ at check-in time ensures $c^T f_i = 5$. We report results computed on a subset of the data: $1074$ users, $8640$ restaurants, and $103$ check-in time buckets. For this setting we study the performance of linear models. The results are averaged over $20$ model instances, where each instance is initialized using a different random seed. We set the embedding size to $k=100$ for all models. An alternating optimization is performed when training all models. Each training iteration represented in the figures corresponds to $100$ optimization steps for items variables, followed by $100$ steps for the users (with the possibility that the optimizer may perform early stopping). $100$ iteration steps are performed for the context variables in the context-aware models, including our models. We perform a warm start for the contextual-constraint-based methods for the initialization of the context vectors. The cost of this operation is similar to one learning iteration, approximately $100$ steps. We report the AUC results computed over the full test set in Fig.~\ref{fig:foursquare_all}. To compute AUC, we need negative examples. We create negatives that depend on the check-in time: if a restaurant check-in is never observed at dinner time, then we assume that this restaurant shouldn't be recommended at dinner time. We then select random restaurants, which we associate with random users that have never visited the same check-in time bucket.
The variance of both weighted and diagonal models is small due to the warm start initialization. As expected, the contextual methods outperform the conventional MF model. We observe that the weighted model (WC-MF) and CAMF-CI give similar results. This is not surprising, as both models have similar expressivity; CAMF-CI considers the feature correlation as a translation, while WC-MF considers the feature as a multiplicative weight. Data enlargement is commonly used with MF models to achieve better performance~\cite{}; therefore, we perform negative sampling based on the observed data, where negatives are sampled according to the check-in times (similar to AUC negative sampling logic we use). As we are using a negative sampling strategy for training that is similar the negative sampling strategy for our evaluation task, we expect very good results for this model configuration. The data enlargement model  results in predictive performance that is similar to context-based models. Our new diagonal transformation method outperforms all of the linear baseline models. We seek to better understand the behaviour of our new model: is this performance improvement explained by an improvement on the similarities of popular contexts, or by better recommendations for rare contexts? We attempt to answer this question by reporting AUC results computed for three different contexts, as shown in Figures~\ref{fig:foursquare_morning}, ~\ref{fig:foursquare_afternoon}, ~\ref{fig:foursquare_night}).  In each figure, we set 3 different check-in time contexts: between $[8am, 9am]$, $[12pm, 1pm]$ and $[10pm, 11pm]$.  AUC is computed using the subset of the training set that contains restaurants observed under the specified constraint. The first two contexts are more rare compared to the third context (evening).
The results for the two rare context confirm that the diagonal model performs better for non-popular contexts, while its performance is similar to the baselines in the case of the popular context. 

\subsection{Low number of overlapping constraints}
We are now interested in observing whether our methods still work when the number of overlapping constraints is low, i.e., when sampling with high probability two constraints, $c_1$ and $c_2$, from the underlying distribution, $c_1^Tc_2 = 0$. In this setting, we cannot perform the same type of warm start as done previously.  In the previous setting, warm start allows the global model to more effectively perform transfer learning using overlapping features.  In the setting described here, the similarities between features are more difficult to recover, since the features almost never overlap, and thus a different warm start scheme is required. In this setting, we leverage the fact that users have historically chosen multiple filters over time, and we assume that the similarity between two contexts is proportional to the number of users that specify both contexts. Similar to the context distribution, the similarity distribution of observed contexts is stable for the train and test sets. 
In this setting, our new models will tend to learn orthogonal transformations, and won't capture similarity based on user behaviour. We add a regularization parameter on $A$ (again we represent $A$ with a $k$-by-$d$ matrix) given by 
\begin{align}
     \sum_{c, c' \in \cC} \frac{\sum_{u: (c,u)\in D_t \text{ and }(c', u)\in D_t} \left(I_k - (A \odot c')^T(A \odot c)\right)^2}{\#\{u: (c,u)\in D_t \text{ and }(c', u)\in D_t\}}
\label{eq:warmstart-data-similarity}
\end{align}
For this model we perform a warm start by initializing $A$ to the minimum of the regularizer \eqref{eq:warmstart-data-similarity}.

\begin{figure}[htbp]
\centering
     \includegraphics[width=0.48\textwidth]{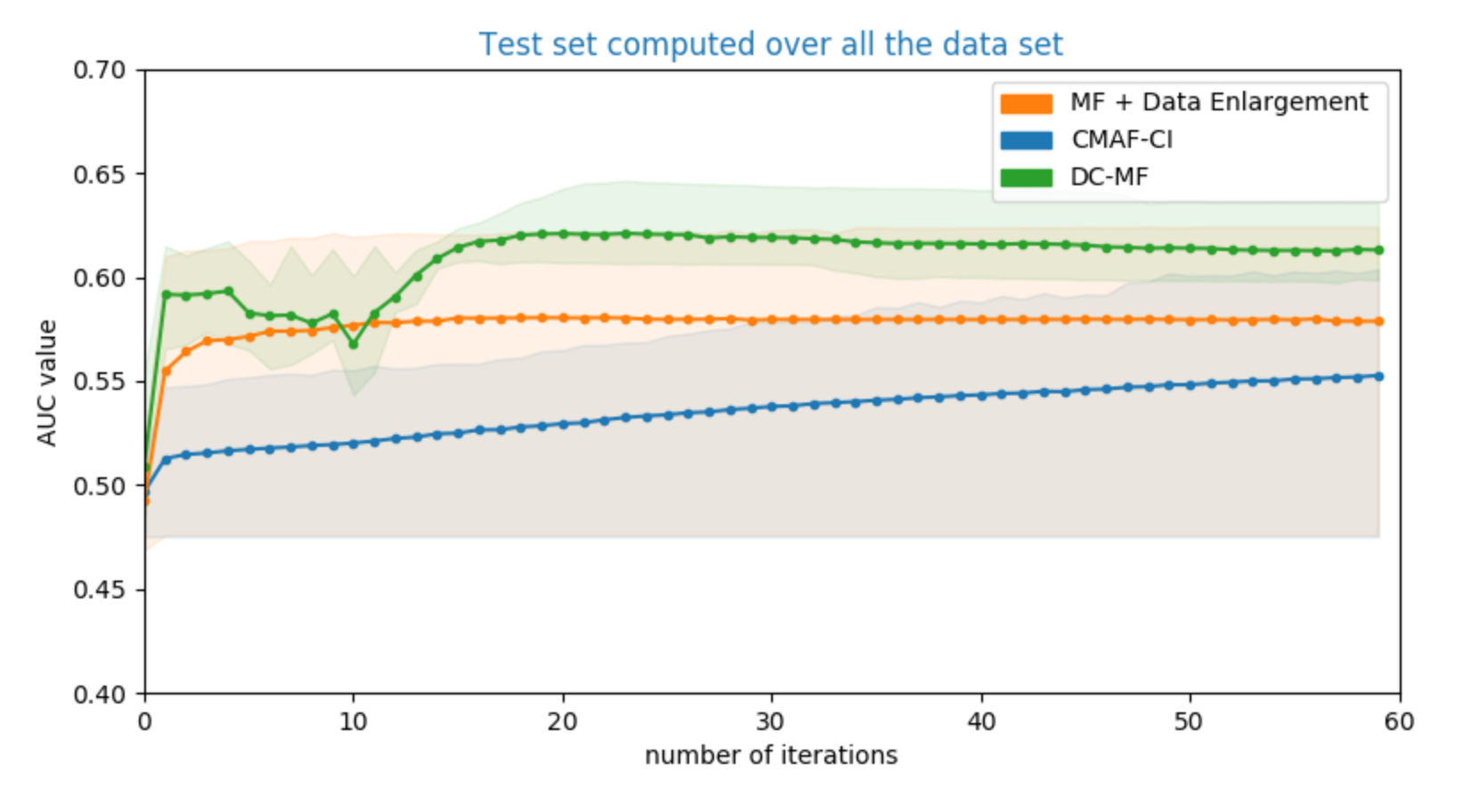}
    \caption{AUC results computed over the global test dataset for the private dataset.}
    \label{fig:brands_all}
\end{figure}

\begin{figure}[htbp]
\centering
     \includegraphics[width=0.48\textwidth]{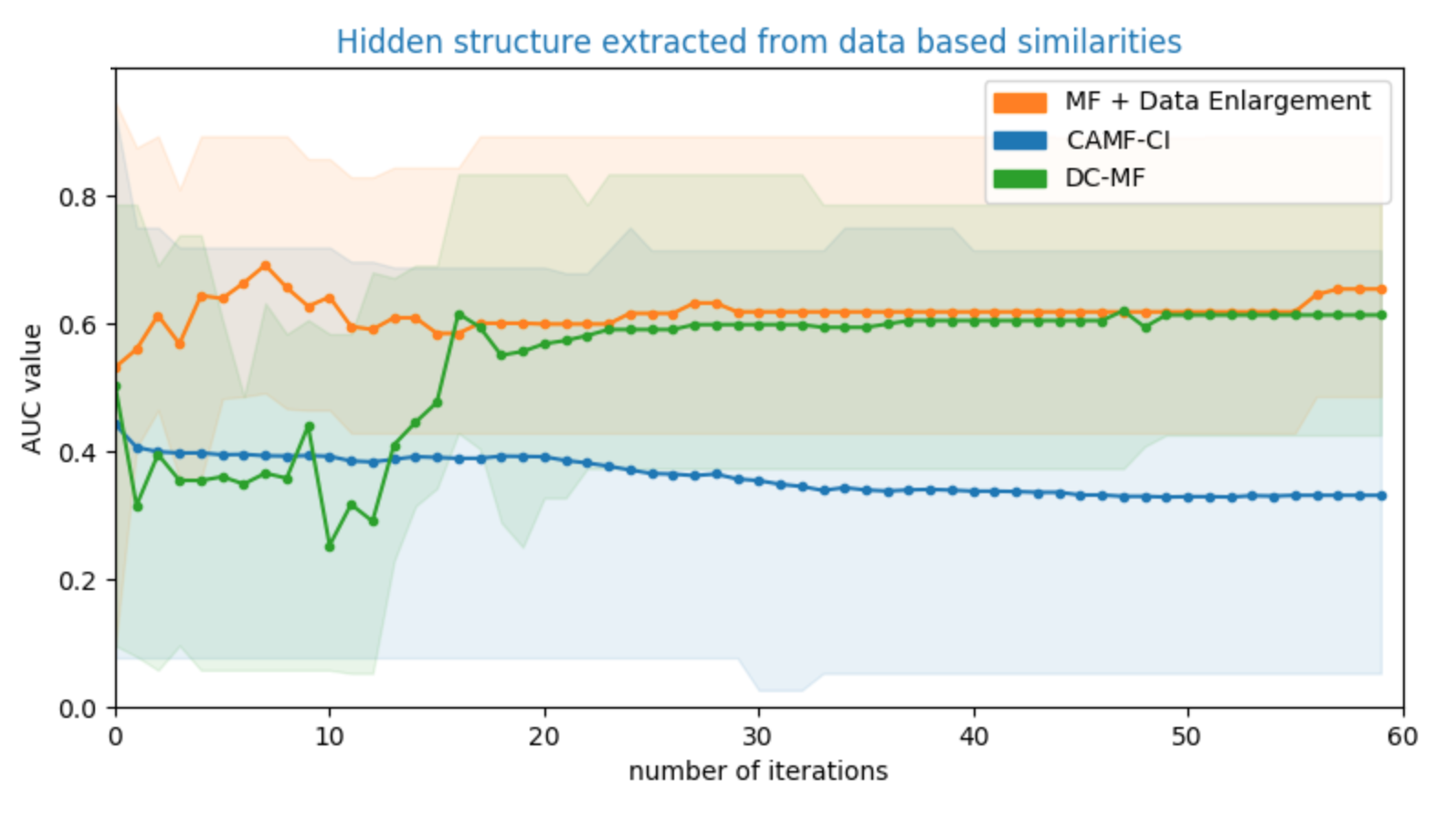}
    \caption{Private dataset: Limits of context-aware models: AUC reported for contextual constraints that specifies multiple brands. The context-aware model can easily overfit on this small dataset when trying to extract contextual similarities. The regularized diagonal-constraint MF model manages to extract similarities from the data, without overfitting.}
    \label{fig:brands_data_sim}
\end{figure}

\begin{figure}[htbp]
\centering
     \includegraphics[width=0.48\textwidth]{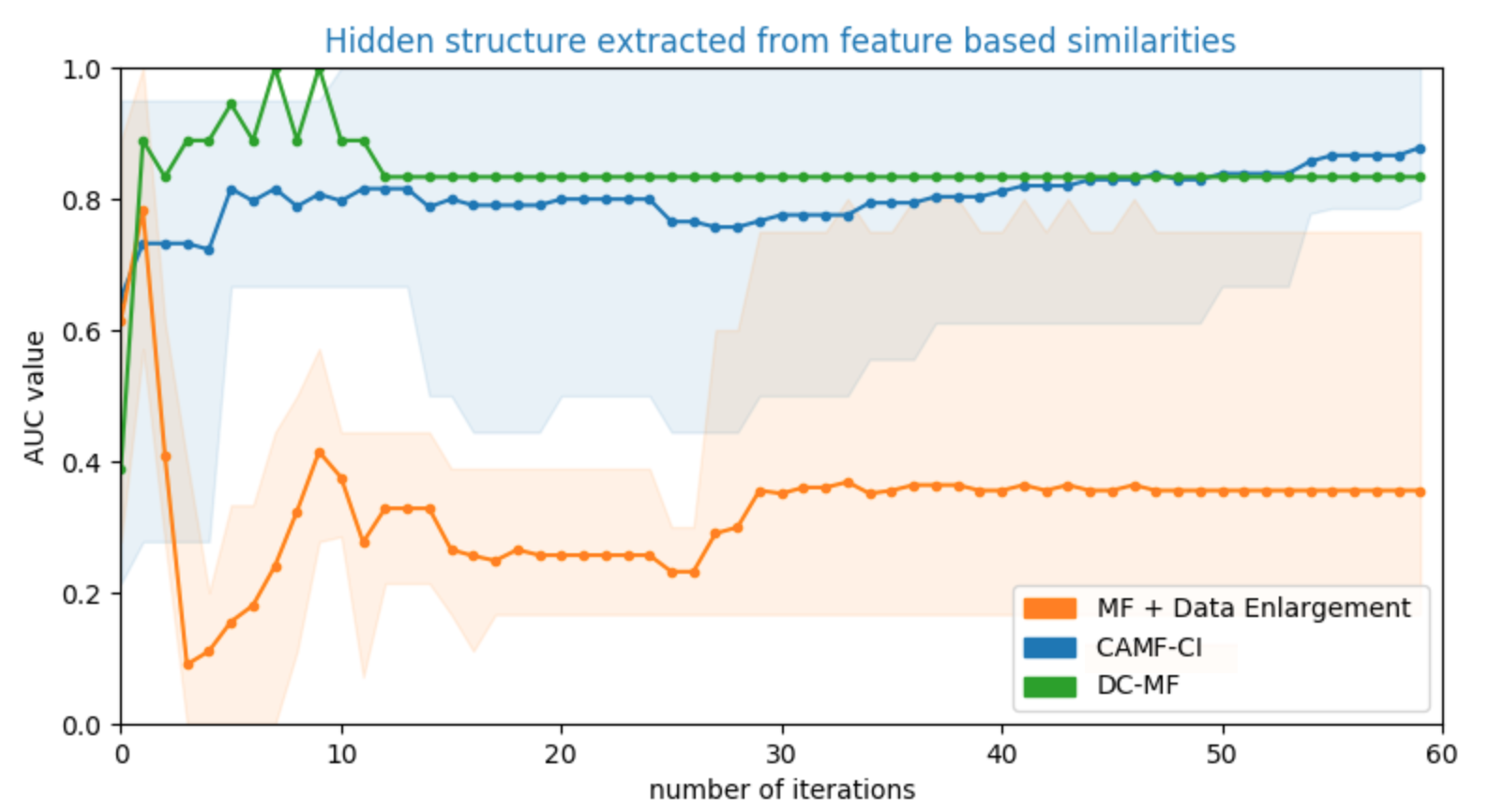}
    \caption{Private dataset: Limits of non-context-aware models: AUC reported for a contextual constraint that sets one brand filter, which is usually set with other brands in other contexts. Compared to the MF model, the context-aware and DC-MF models benefit from refining recommendations based on feature similarities.}
    \label{fig:brands_feature_siml}
\end{figure}

\paragraph{Private Criteo dataset: product recommendation under brands constraints}
This private dataset has been extracted from a retailer dataset. The recommendation is constrained by filters set by the user.  These filters are strict; nothing is displayed if a filter is not satisfied. For this dataset we are confident about the constraints, which represent brand filters. One user can set multiple filters. We observe that our dataset is very sparse, since filters are rarely set together. The reward is more explicit, as we collect information about clicks and non-clicks. We run experiments on a subset of the observed data: $m=11,655$ users, $n=2564$ items, and constraint features that correspond to $d=363$ brands. We set the embedding size to $k=100$ for all models. The positives and negative samples are used to train all models. Due to the imbalance of the observed clicked and non-clicked data, we re-weight the positives and and negatives in order to provide reasonable performance for the MF model. We perform one warm start iteration for our new model. We train the models with $10$ different random seeds, and report the results in  Fig.~\ref{fig:brands_all}.
Overall, the three models gives close to average performance results. DC-MF still gives slightly better results, as expected. With the regularization parameter the diagonal constrained model finds a better trade-off between the use of feature-based similarities and user behaviour similarities. Fig.~\ref{fig:brands_data_sim} and \ref{fig:brands_feature_siml} shown this trade-off. In Fig.\ref{fig:brands_feature_siml},
DC-MF provides performance similar to the context aware model, while MF fails to learn non-contextual similarities. In Fig.~\ref{fig:brands_data_sim}, DC-MF maintains good performance, while the context-aware model fails. DC-MF is thus better to use for recommendation under the brand constraints in this setting, as we ensure lower-variance recommendations across the brands. These two figures also show a limitation of the diagonal model: when the overlap between the constraints is low, the diagonal model only outperforms the CAMF-CI model to a small extent.

\subsection{Constraints with multiple features and avoiding the folding effect}
\label{ssec:nn_experiments}
In this setting we observe users that interact only with items in a restricted subset of contexts, where these subsets are disjoint: If we denote $C_1$ and $C_2$ as the two disjoint subsets of contexts, then $\forall c_1 \in C_1$ and $\forall c_2 \in C_2$,  $c_1^Tc_2  = 0$, and $\not \exists u \in [m] | (u,c_1), (u,c_2) \in D_t$. This typically corresponds to the situation where a sub-population of users does not see a large part of the set of items because of constraint restrictions, e.g., horror movies cannot be recommended to kids, and therefore kids won't watch or rate these movies. This can result in bad behaviour, where the recommender system may accidentally learn a high similarity between kids and horror movies, due to lack of information. This folding effect was introduced in~\cite{xin2017folding}, where the model can learn similar vector representations for users and items that belong to disjoint sets of constraints. In this scenario, the model learns similarities that don't exist in the data. The true similarity between the two disjoint contexts, $c_1$ and $c_2$, is actually zero, since we never observe users setting both contexts. The spurious recommendation phenomenon induced by folding effects is observed when filtering is done after item scoring. Therefore, the probability of seeing the $k$ best products satisfying a particular context in the tail of the predictive distribution is high.

In scenarios which can result in folding, it is crucial to have additional metadata descriptors for users, items, and/or the constraints, in order to generalize well for cases that are not supported by the feedback distribution, as described above. Therefore, we use neural network based approaches for this experiment, since neural networks allow us to easily accommodate metadata into the model. The baseline is the NN-MF model described in~\eqref{eq:nnmf}, where the network takes as input both the user and user features for $U$, and the item and item features for $P$. We also introduce a slightly modified version of NC-MF: instead of using MF embeddings for $U$ and $P$ (see~\eqref{eq:loss_model}), we use the neural network to learn user and items embeddings that take into account metadata descriptors, in order to provide a fair comparison with NN-MF. We refer to this modified version of NC-MF as NC-NN-MF.

\paragraph{MovieLens dataset}
The MovieLens 100K dataset consists of $m=943$ users and $n=1682$ movies. Each user has rated at least $20$ movies. 
This dataset also contains information about the user, such as age, and movie categories. We join the logs in the dataset in order to link user features, item features, and ratings.  In this dataset we expect that kids cannot watch horror or "thriller" movies, and therefore cannot rate those movies. We expect that the learned models are capable of detecting that recommending thriller or horror movies to kids is considered to be a bad recommendation.
To better study folding effects, we select a training set where the observed ratings on horror movies are disjoint from the other movie categories. We then select two test sets that are subject to folding effects:
\begin{itemize}
    \item Horror movies: user and items in this category are disjoint with the other categories.
    \item Thriller movies: user and item in this category are disjoint with the other categories for kids only. 
\end{itemize}
We compute the AUC over both test sets by adding explicit negatives for kids to only the test set. We would like the model to recommend $-1$ for kids (age $< 14$), and a positive value for adults (age $>=14$). The models are run using $10$ random seeds.

\begin{figure}[htbp]
\centering
     \includegraphics[width=0.48\textwidth]{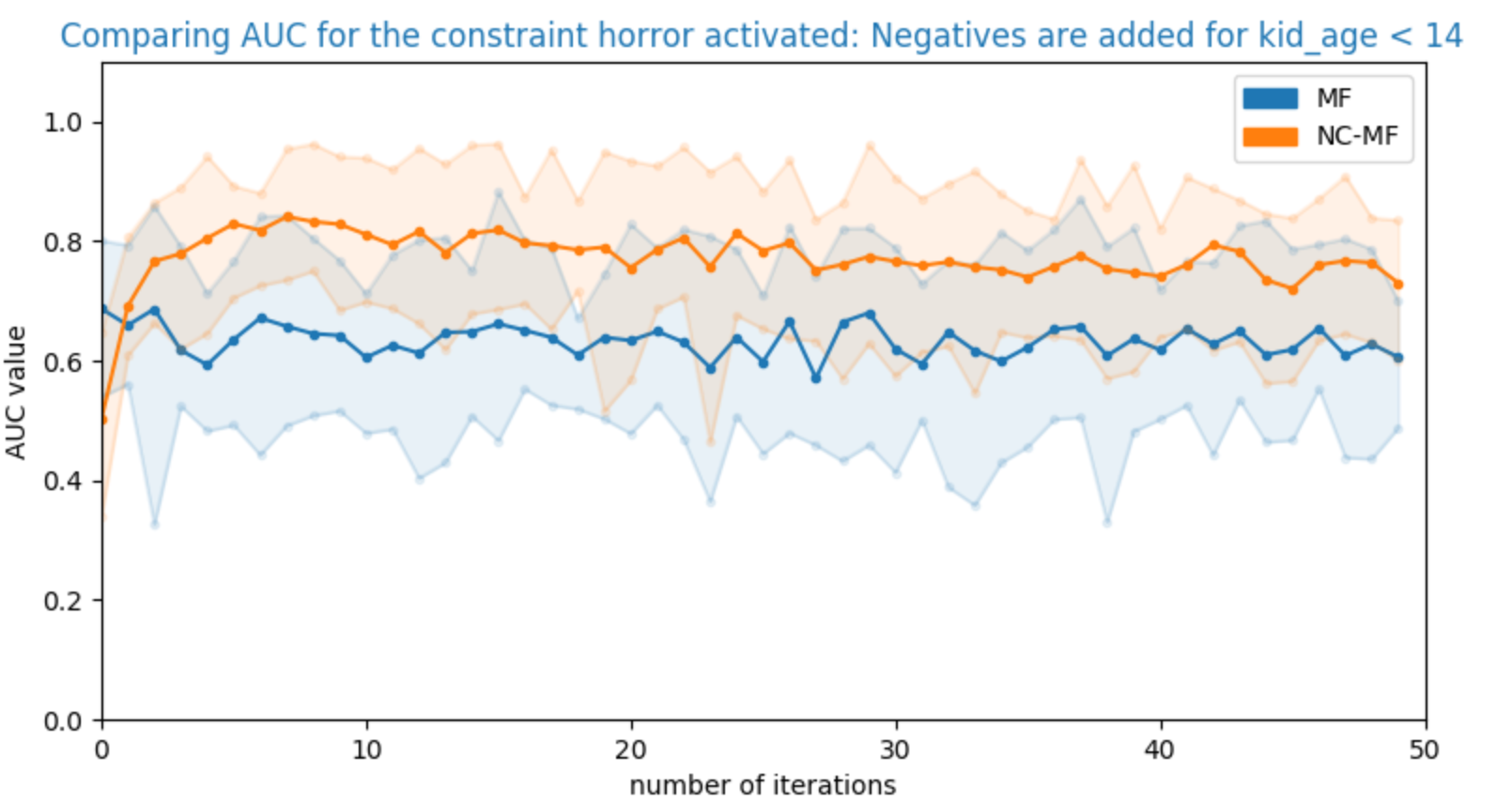}
    \caption{MovieLens: AUC on the horror movie test set. The models are learned based on user ids and item ids.  For our new models we add the three context features: a binary indicator for a thriller movie, a binary indicator for a horror movie, and the age of the user.}
    \label{fig:mv_horror}
\end{figure}

\begin{figure}[htbp]
\centering
     \includegraphics[width=0.48\textwidth]{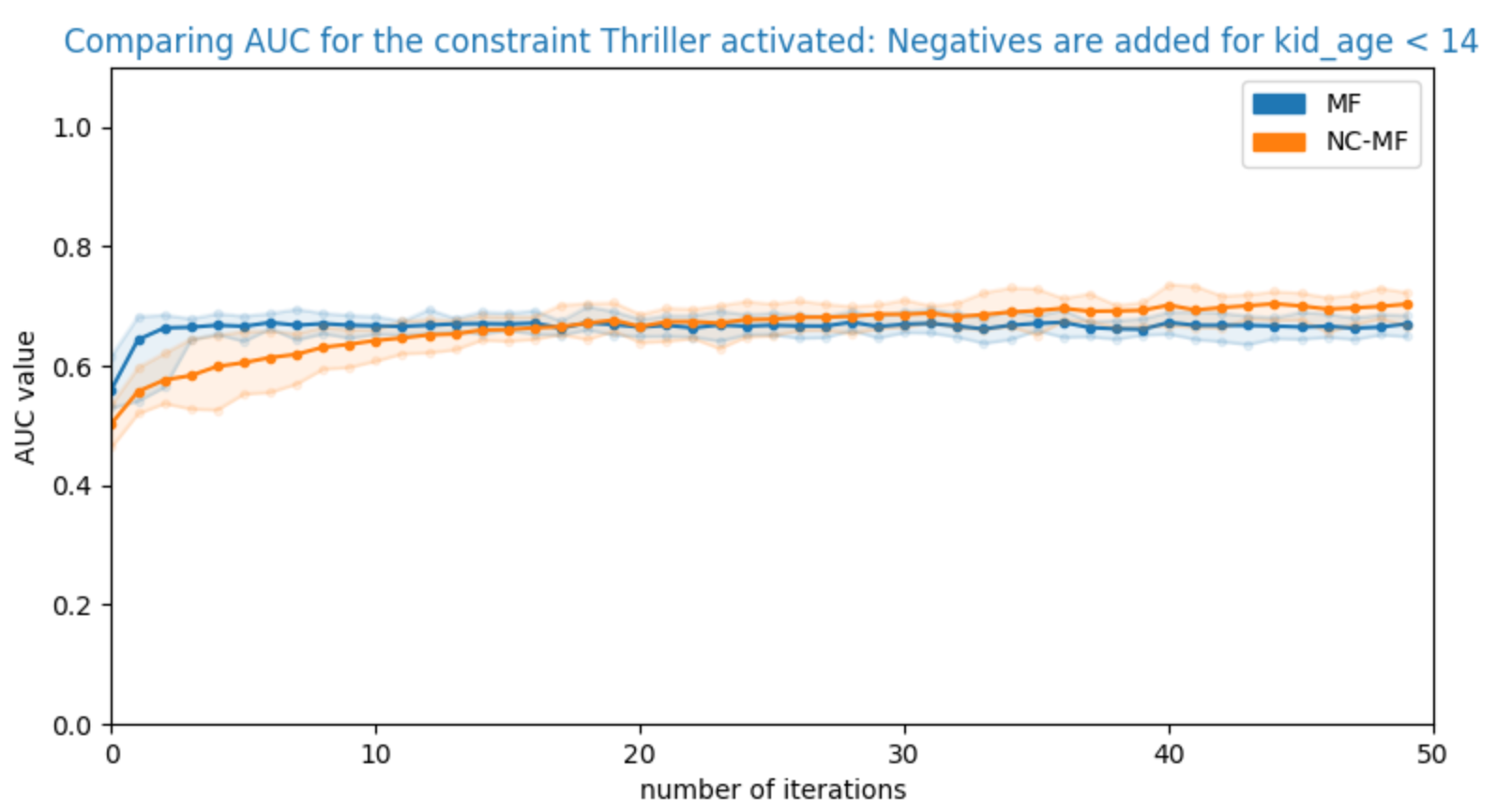}
    \caption{MovieLens: AUC on the thriller movie test set. The models are learned based on user ids and item ids.  For our new models we add the three context features: a binary indicator for a thriller movie, a binary indicator for a horror movie, and the age of the user.}
    \label{fig:mv_thriller}
\end{figure}

We compare the results of the NC-MF model to a normal MF model. 
As the new models will learn orthogonal representations for the two different disjoint contexts, folding effects are reduced. In Fig.~\ref{fig:mv_horror}, we observe a lower risk of a low AUC value ($<= 0.5$) for horror movies. Both MF and NC-MF perform well on the thriller dataset, shown in Fig.~\ref{fig:mv_thriller}, as it is easy to detect that the movies rated by kids are different from those rated by adults.

\begin{figure}[htbp]
\centering
     \includegraphics[width=0.48\textwidth]{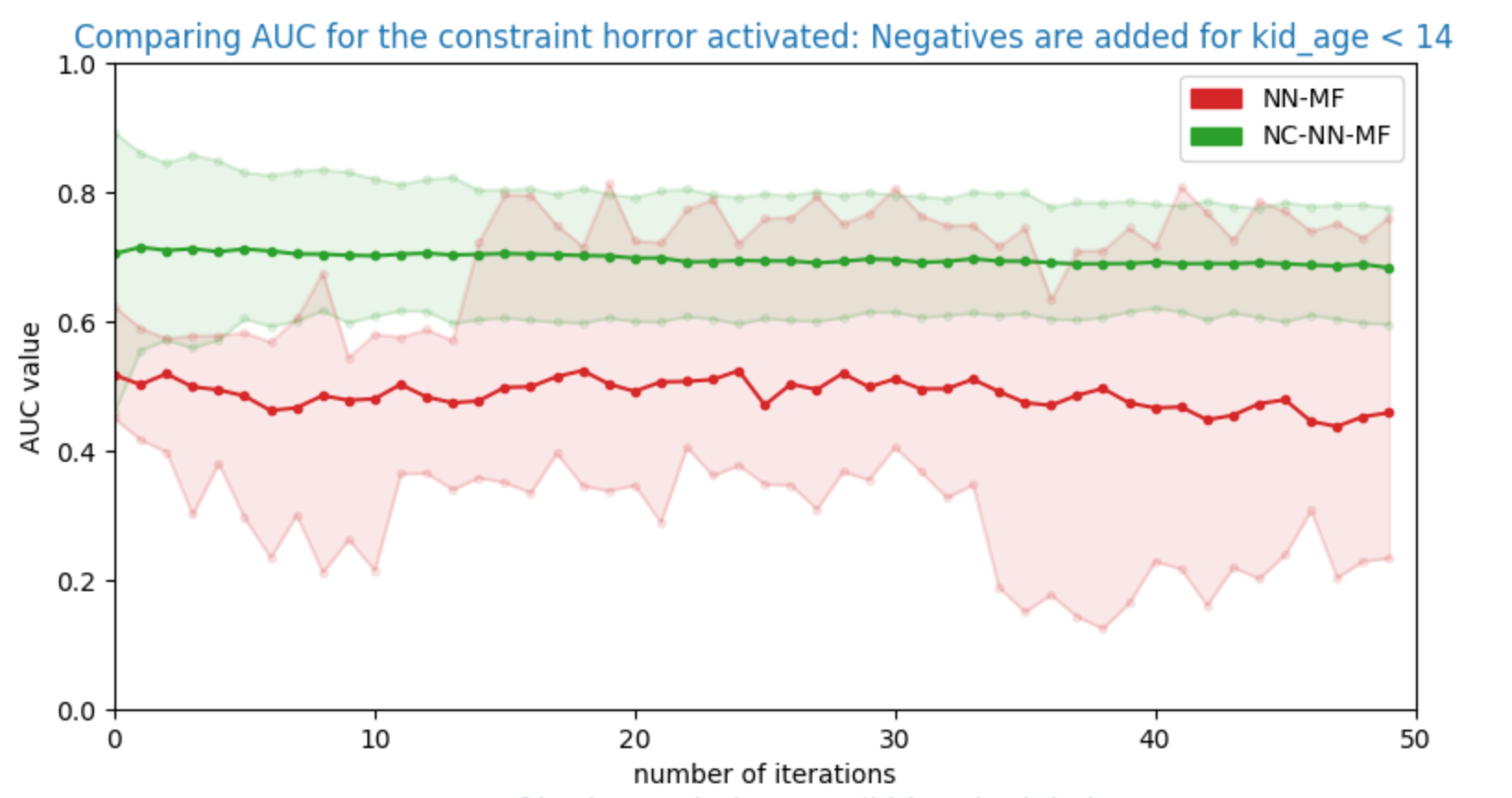}
    \caption{MovieLens: AUC on the horror movie test set for the NN-MF models. The neural net takes as input user and item features, along with the user and item ids. For the context based model (NC-NN-MF), we use the three contextual features, and we remove user age from the neural net input, so as prevent use of this feature twice.}
    \label{fig:nmf-horror}
\end{figure}

\begin{figure}[htbp]
\centering
     \includegraphics[width=0.48\textwidth]{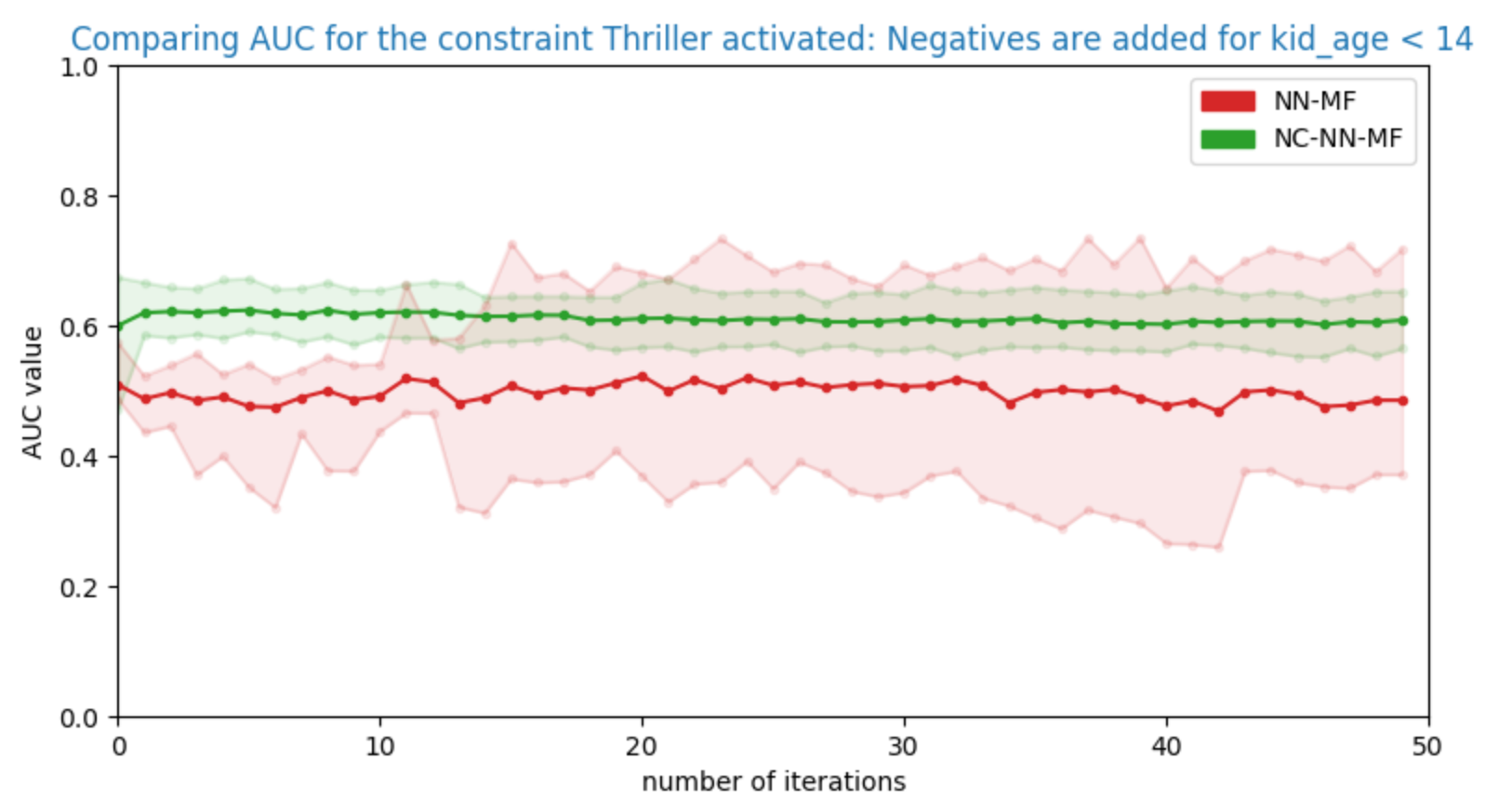}
    \caption{MovieLens: AUC on the thriller movie test set for the NN-MF models. The neural net takes as input user and item features, along with the user and item ids. For the context based model (NC-NN-MF), we use the three contextual features, and we remove user age from the neural net input, so as prevent use of this feature twice.}
    \label{fig:nmf-thriller}
\end{figure}

Although we expect NN-MF to provide lower performance due to the size and sparsity of the dataset, we report results comparing NC-NN-MF compared to NN-MF. As described above, NC-NN-MF is one of our new constraint models, where we apply a neural constraint transformation on top of a neural net matrix factorization. As in NC-NN-MF, we explicitly define the hidden structure to learn; the model easily captures the relationship between the user age and item category. In Fig.~\ref{fig:nmf-horror}, we see that NN-MF is capable of extracting the correct hidden structure for some random seeds, but fails for other seeds, leading to high variance in the predictions. By limiting the combination of features in NC-NN-MF so that we explicitly provide the features that defines the neural constraint, we restrict the expressivity of the model. Thus, for non-disjoint categories of items, such as thriller movies, we for some seeds NN-MF outperforms NC-NN-MF, as shown in Fig.~\ref{fig:nmf-thriller}.

%% file: conclusion.tex
\section{Conclusion and Future Work}
\label{sec:conclusion}
Real-word recommender system are often constrained by partially observable or recommendable item-user pairs. Any additional information about the user's interest may help to refine the recommendation and improve generalization on unobserved user-item pairs. For some recommender system settings, the user can interact with the system to refine recommendations by using filters. This information is rarely incorporated when training the recommendation model. Due to the sparsity of the observed data and the combinatorial nature of the set of constraints, incorporating contextual constraint information can be a difficult task, and for some settings non-contextual models better recommendations than context-aware models. We propose a new framework that describes recommendation under contextual constraints, where the observations are very sparse. We then present new methods that incorporate the new information provided by these constraints in different ways. Our new models learn a constraint representation as a similarity measure between user and item embeddings. This representation allows to increase the expressivity of the model without substantially increasing the number of model parameters. We perform experiments on three datasets that involve several challenging constraint-based settings. We describe an adaptation of the model to solve each of these tasks by using a warm start and/or regularization scheme appropriate to each setting. Our new model achieves better performance than the baseline models we consider for each setting. A promising direction for future work would be to incorporate recently proposed deep models for learning on sets~\cite{NIPS2017_6931}, which would allow us to have a more elaborate representation of the constraints.